\documentclass{ceurart}

\usepackage[T1]{fontenc}
\usepackage{mathptmx}
\usepackage{soul}\setuldepth{article}
\usepackage{booktabs}
\usepackage{algorithm}
\usepackage{algorithmic}
\usepackage{amsmath}
\usepackage{subcaption}
\usepackage{graphicx}
\usepackage{breqn}
\usepackage{url}
\usepackage[Q=yes]{examplep}
\usepackage[inline]{enumitem}
\usepackage{listings}
\usepackage{verbatim}
\usepackage{todonotes}
\usepackage{longtable}
\usepackage{enumitem}
\usepackage{pgfplots}
\usepackage{pgfplotstable}

\newtheorem{example}{Example}

\def\hb{\hbox to 11.5 cm{}}

\begin{document}

\copyrightyear{2025}
\copyrightclause{Copyright for this paper by its authors. Use permitted under Creative Commons License Attribution 4.0 International (CC BY 4.0).}

\conference{}

\title{A SHACL-based Data Consistency Solution for Contract Compliance Verification (Extended Paper)}

\author[1,2]{Robert David}[orcid=0000-0002-3244-5341,
email=robert.david@graphwise.ai
]

\author[1,3]{Albin Ahmeti}[
orcid=0000-0001-8766-4069,
email=albin.ahmeti@graphwise.ai
]

\author[4]{Geni Bushati}[
email=geni.bushati@sti2.at
]

\author[4,5]{Amar Tauqeer}[
orcid=0000-0002-3345-387X,
email=amar.tauqeer@wur.nl
]

\author[5]{Anna Fensel}[
orcid=0000-0002-1391-7104,
email=anna.fensel@wur.nl
]

\address[1]{Semantic Web Company GmbH, Austria}
\address[2]{Vienna University of Economics and Business, Austria}
\address[3]{Vienna University of Technology, Austria}
\address[4]{Semantic Technology Institute, Department of Computer Science, Universität Innsbruck, Austria}
\address[5]{Artificial Intelligence Chair Group, Wageningen University \& Research, The Netherlands}

\begin{abstract}
In recent years, there have been many developments for GDPR-compliant data access and sharing based on consent. For more complex data sharing scenarios, where consent might not be sufficient, many parties rely on contracts. Before a contract is signed, it must undergo the process of contract negotiation within the contract lifecycle, which consists of negotiating the obligations associated with the contract. Contract compliance verification (CCV) provides a means to verify whether a contract is GDPR-compliant, i.e., adheres to legal obligations and there are no violations. 
The rise of knowledge graph (KG) adoption, enabling semantic interoperability using well-defined semantics, allows CCV to be applied on KGs. 
In the scenario of different participants negotiating obligations, there is a need for data consistency to ensure that CCV is done correctly. 
Recent work introduced the automated contracting tool (ACT), a KG-based and ODRL-employing tool for GDPR CCV, which was developed in the Horizon 2020 project smashHit (\url{https://smashhit.eu}). 
Although the tool reports violations with respect to obligations, it had limitations in verifying and ensuring compliance, as it did not use an interoperable semantic formalism, such as SHACL, and did not support users in resolving data inconsistencies. 
In this work, we propose a novel approach to overcome these limitations of ACT. 
We semi-automatically resolve CCV inconsistencies by providing repair strategies, which automatically propose (optimal) solutions to the user to re-establish data consistency and thereby support them in
managing GDPR-compliant contract lifecycle data. 
We have implemented the approach, integrated it into ACT and tested its correctness and performance against basic CCV consistency requirements.
\end{abstract}

\begin{keywords}
  Privacy protection \sep
  GDPR \sep
  Contract compliance verification \sep
  Data consistency \sep
  Constraint languages \sep
  SHACL \sep
  Knowledge graphs \sep
  Logic programming \sep
  Answer set programming
\end{keywords}

\maketitle

\section{Introduction} \label{sec:Introduction}
%
%
%
%
%

General Data Protection Regulation (GDPR) \cite{gdpr} came into effect in the EU on 25 May 2018, defining strict requirements for data processing, management, and storage~\cite{Li2018DSAPDS}. 
One of the legal bases that must be satisfied for GDPR is \emph{informed consent}. 
Collecting informed consent from a \emph{data subject} (i.e., an identifiable natural person) (Art. 4 (1)) before sharing the actual data is one of the most common steps done by various organisations nowadays. 
There are certain cases where consent might not be enough, for example, in cases where parties need to discuss the details of the \emph{contract} for data sharing. This is typical for online services that comprise Business-to-Business (B2B) and Business-to-Consumer (B2C) data sharing, where a contract specifies the terms and obligations specifying what each party's responsibilities are and what they are allowed to do. Referring to~\cite{TauqeerKCAF22}, the main differences between consent and contracts are as follows: \textit{``(i) consent can be revoked at any time, while a contract cannot be terminated before its minimum duration; (ii) consent has predefined clauses, while contract clauses can be negotiated until an agreement is reached between all involved parties''}.
The latter item is called \emph{contract negotiation} and is part of the contract lifecycle. The contract lifecycle comprises all the phases that are involved in establishing a contract (and the associated longevity) between parties in the following order: negotiation, signing, execution, auditing, and termination/renewal phase. In this context, \emph{Contract Compliance Verification (CCV)} is about auditing and provides a means to verify whether a contract is GDPR compliant, that is, to check and report violations in case a party did not adhere to regulations, e.g., for not fulfilling an obligation that is mentioned in a term (clause). 
The requirements pertaining to contract negotiation and CCV call for a more scalable, loosely coupled architecture, like the ones typically seen in publish-subscribe distributed systems~\cite{Kleppmann2016}, in which the communication is done via a set of events, eventually leading to an agreement of terms between parties. 
As a result, consistency issues can arise in this type of architecture, requiring methods to employ integrity constraints to ensure a consistent contract lifecycle. 

Recently, we have seen approaches for digital contracting to tackle the CCV challenge based on knowledge graphs (KGs)
~\cite{TauqeerKCAF22} as a means to improve interoperability, interpretation, and contextualization of data, employing precise semantics through \emph{ontologies} and \emph{controlled vocabularies}. 
The automated contracting tool (ACT)~\cite{Tauqeer2024-rp}, which uses KGs with standardised semantic descriptions for contract lifecycle data, has been tested and used in two data sharing scenarios, namely for car insurances and for smart cities. Especially in the latter case, large amounts of data pertaining to contracts are shared, calling for a more scalable, trusted, and secure platform for data sharing. 
%
%
We define our challenges as:
\vspace*{1pt}
\begin{enumerate}[label=(\roman*)]
    \item Identifying and reporting CCV violations by defining and integrating constraints for ACT using the Shapes Constraint Language SHACL \cite{shacl}.
    \item Fixing CCV violations, i.e. 
    semi-automatically repairing identified violations while at the same time ensuring the semantics of the contract.
\end{enumerate}
To cope with these challenges, we first establish integrity constraints using SHACL. 
In the context of the ACT tool, it may occur that CCV violations are reported based on inconsistencies present in received data. 
For such cases of CCV violations, we employ SHACL validation as a declarative approach to identify inconsistencies, and \emph{SHACL repairs} \cite{10.1007/978-3-031-19433-7_22} to semi-automatically correct identified inconsistencies with a minimum of user intervention. 
This is achieved by defining and implementing repair strategies for inconsistent data reported by CCV. These repair strategies i) extend our previous work on SHACL repairs \cite{10.1007/978-3-031-19433-7_22} to provide use case specific repair optimizations and ii) integrate with our ACT tool for users to semi-automatically fix inconsistent data reported by CCV. 

Our paper is structured as follows: 
we discuss related work in the area of automatic contracting tools in Section \ref{sec:RelatedWork}, and preliminaries for constraint validation and repairs in Section~\ref{sec:preliminaries}. Then we present the architecture and the semantic model for the contract lifecycle in Section~\ref{sec:Approach}.
The main part, Section~\ref{sec:ccv-constraints}, describes our application for consistency and repair based on SHACL constraints, followed by the evaluation we performed in Section~\ref{sec:evaluation}. 
Finally, we conclude by outlining future work in Section~\ref{sec:Conclusion}. 

\section{Related Work} \label{sec:RelatedWork}
%
%
%
%
%
Manually sorting through stacks of paper contracts or even digital contracts in a disorganised folder can be time-consuming and overwhelming. Visualisation of contracts through automated software tools allows for greater efficiency and organisation. Moreover, these automated tools for contract management can help identify potential issues or problems within contracts. More precisely, certain clauses or terms may be missing or contradictory, thus leading to legal issues down the line. Automated Contracting Tools offer features that can scan contracts for these types of issues and flag them for review by the involved parties, thereby reducing the risk of non-compliance and potential penalties. 
There are several systems developed to manage contracts consistently between participants, like described in \cite{brown2018corda} and \cite{guo2021blockchain}. 
Our previous work stemming from the smashHit project, introduced the data sharing and contract management ACT tool that was applied to use cases in the car insurance and smart city domains. ACT invokes a CCV service on-demand or periodically in order to ensure GDPR compliance is met and consequently there are no violations in respect to contracts' obligations. 

Web services are a dominant approach in the domain of integration of applications, driving forward the current IT industry by providing solutions for businesses and services. Often, these types of systems need to work together in an open environment, thus making the compliance outcome less predictable. 
Related work for automated contract management
can be found in \cite{10.5555/1402383.1402424}, Jang et al. \cite{10.1145/2629630} and \cite{molina2011model}.

Generally, work on repairs for Description Logic ontologies is related to our approach. 
A recent work \cite{10.1007/978-3-031-06981-9_8} addresses optimal repairs in the scenario where the schema (TBox) is assumed to be correct, while the data (ABox) needs to be repaired. 
This is in the same spirit of our approach to define a schema via SHACL and then repairing the CCV (ABox) data accordingly. 

Regulation compliance checking is shown in \cite {bouzidi2012semantic} using ontologies and SPARQL. 
Implementations for automated compliance checking for RDF data are presented in \cite{10.1093/logcom/exad034}. 
They investigate how logic programming can be leveraged for RDF constraint checking and present how the DLV2 reasoner can be directly integrated with RDF data for Answer Set Programming.
Compared to these related works, we go a step further and present a novel SHACL-based application for automated repairs to ensure compliance, and show its feasibility by implementing in the ACT tool. 

\section{Preliminaries}
\label{sec:preliminaries}
%
%
%
%
%
Our recent publications \cite{10.1007/978-3-031-19433-7_22} introduced a novel approach to repair SHACL constraint violations by modifying the data to conform to the defined constraints. 
The SHACL repair approach is implemented as a repair program that can deduce repairs as cardinality minimal data modifications.
It will generate consistent data regarding the shape constraints when applied to a data graph. The repair program implementation is available on GitHub\footnote{\url{https://github.com/robert-david/shacl-repairs}}. 

The repairs come in sets of additions and deletions \cite{KR2021-2} of triples determined by the repair program for an input data graph and a set of SHACL shapes to validate against. Adding the additions to the data and removing the deletions from the data will result in a new data graph that is consistent with the shape constraints. 
The implementation of the repair program is done by a Java program, which  reads the SHACL shapes and the data graph and generates an answer set program (see Answer Set Programming \cite{EiterIK09}) that can be processed by the clingo \cite{gebser_kaminski_kaufmann_schaub_2019} solver. 
Answer set programming (ASP) provides minimal models as solutions to a given program. By building on this minimality of ASP models, the repair program provides minimal repairs as part of these minimal models.
Also, the ASP program can even resolve potentially conflicting constraints, eventually stabilizing into a stable minimal model. 
However, it is possible that there are multiple minimal models with different repairs returned for a given program. We illustrate this situation with the following example. 
\begin{example}
Consider the following SHACL shape.
\begin{lstlisting}
:ContractShape a sh:NodeShape;
  sh:targetClass fibo:Contract;
  sh:property [ sh:path smashHitCore:hasContractStatus; 
  sh:maxCount 1; ] .
\end{lstlisting}
The \emph{ContractShape} defines a constraint for a maximum of one triple (atom) of property \emph{hasContractStatus}.
Consider the following data graph.
\begin{lstlisting}
:contb2b a fibo:Contract;
  smashHitCore:hasContractStatus smashHitCore:statusPending, 
  smashHitCore:statusFulfilled.
\end{lstlisting}
The SHACL repair program will propose two different minimal models to repair the data graph to satisfy the shape constraint. These models contain the deletion sets $D_1$ and $D_2$ respectively.
\small
\small
\begin{align*}
    D_1 = \{&hasContractStatus(contb2b,statusPending)\} \\
    D_2 = \{&hasContractStatus(contb2b,statusFulfilled)\}
\end{align*}
\normalsize
Picking one of these sets and applying it to the data graph for deletion will result in a data graph which conforms to the maximum cardinality constraint defined in the shape. However, the choice for one of these two repair models is not made by the repair program. In the next section we see how a user can choose one of the repair options in ACT frontend. 
\end{example} 

\section{CCV Architecture}\label{sec:Approach}
%
This section presents the CCV architecture. We first provide a brief overview of the contract lifecycle given that it sets the context, followed by the system architecture, the semantic model and the ACT application for managing contracts. 

\subsection{Contract Lifecycle}
\par The contracting process begins with a draft, followed by the negotiation step where contractors review it. 
In this step, terms are negotiated as well, where each term may contain a set of obligations for the contract.
The next step ``Signing'' is executed when contractors reach an agreement on terms and clauses, which is then followed by an ``Execution'' step, where obligations start to get fulfilled.
The next step is ``Auditing and Controlling'', where CCV checks are performed to verify the validity of a contract. This step is the focus of our work.
Finally, the ``Termination/Renewal'' step ensures that the contract is either terminated or renewed, thus concluding the contract lifecycle. 
%

%
%
%

\subsection{System Architecture and Workflow}\label{sec:Implementation}
\begin{figure}[!ht]
        \includegraphics[scale=0.15]{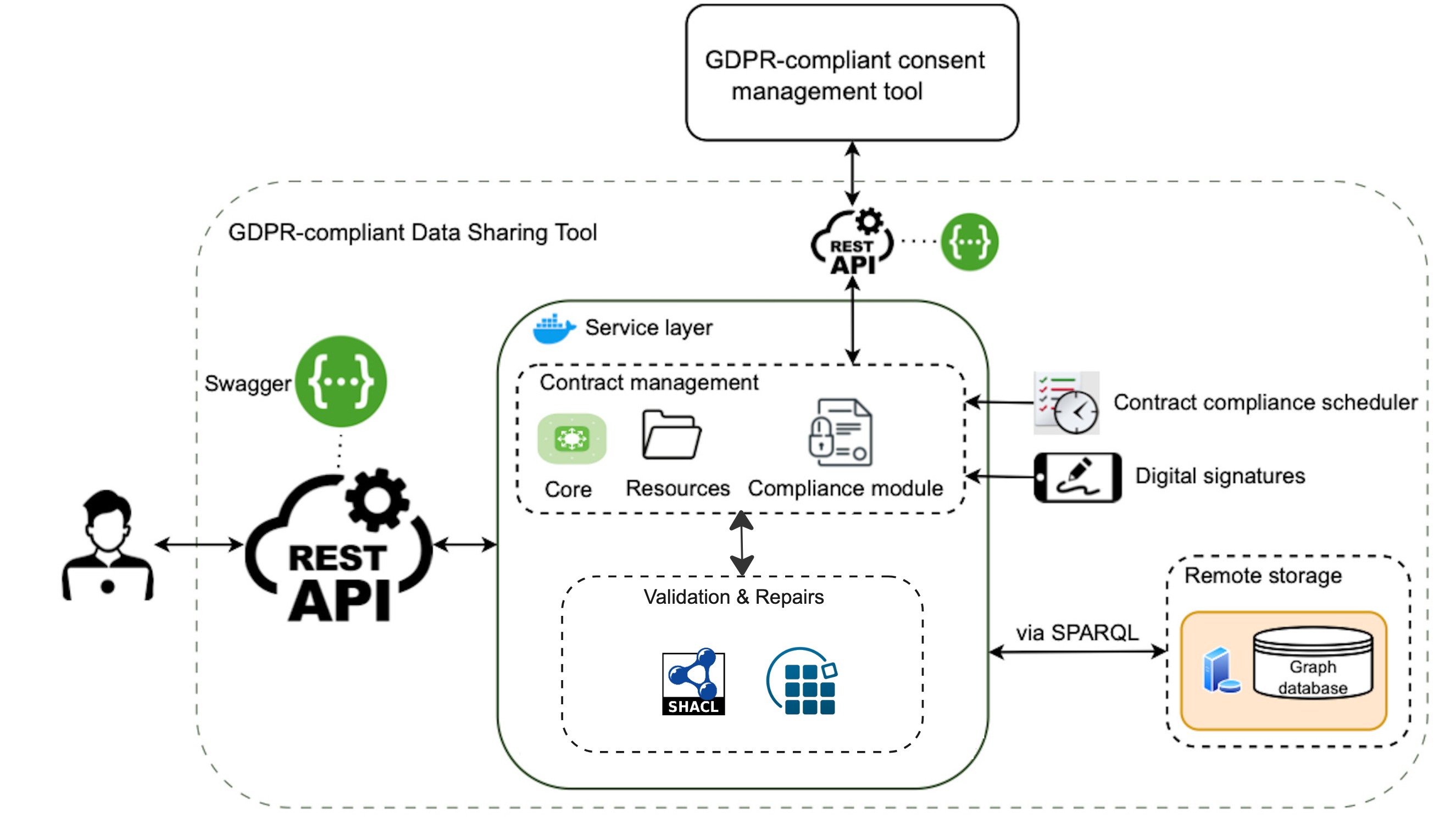}
    \centering
    \caption{The extended architecture with SHACL validation and repairs.} \label{fig:architecture}
\end{figure}

In this work, we focus on consistency for data processing for our previously developed architecture \cite{TauqeerKCAF22}. Consistency is implemented as a separate new component extending the architecture by introducing SHACL validation and repair strategies (cf.\ Fig.\ \ref{fig:architecture}).

With this new solution, we moved away from checking consistency in the application code, because this was not flexible and hard to maintain compared to using SHACL as a declarative standards-based approach.
In the contract lifecycle, when creating and changing a contract, or obligation, respectively, in each of the steps in CCV the implementation of validation is done via a SHACL processor. The validation process checks the conformance with respect to obligations and \emph{consistency requirements}, e.g., checks if the status of a contract is consistent with the obligation states. 

\subsection{Semantic Modeling} \label{sec:SemanticModelling}
The data for contracts and associated elements is described using the Contract Ontology (smashHitCore) as described in \cite{TauqeerKCAF22}, by using classes such as \emph{Contract} and \emph{Obligation}. 
The Contract Ontology reuses the Financial Industry Business Ontology (FIBO), which can be used to semantically describe elements relevant for the contract lifecycle\footnote{\url{https://spec.edmcouncil.org/fibo/ontology/FND/Agreements/Contracts/}}.
In this work, we focus on the \emph{fibo:Contract} class, which has its status represented using the property \emph{smashHitCore:hasContractStatus}, and the \emph{smashHitCore:Obligation} class, in which a contract is associated with via the \emph{smashHitCore:hasObligations} property and its state is represented using the property \emph{smashHitCore:hasState}. 
\begin{figure}[!ht]
        \includegraphics[scale=0.45]{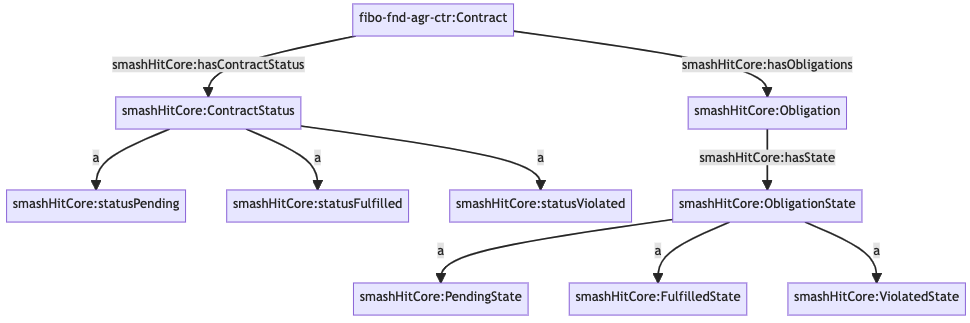}
    \centering
    \caption{Semantic Representation of Contracts and Obligations. Only the relevant part of the Contract Ontology is shown.} \label{fig:contract-ontology}
\end{figure} 
%
%
\begin{example}
\label{ex:contract-graph}
In the following we provide an example of a contract, its status, its obligations and their respective states. 
\begin{lstlisting}
:contb2 a fibo:Contract; 
  smashHitCore:hasContractStatus smashHitCore:statusPending ;
  smashHitCore:hasObligations :ob_1, :ob_2 .
:ob_1 a smashHitCore:Obligation; smashHitCore:hasState smashHitCore:PendingState .
:ob_2 a smashHitCore:Obligation; smashHitCore:hasState smashHitCore:FulfilledState .
\end{lstlisting}
The contract has a pending state since there is one obligation state that is pending. Otherwise, if the obligation state eventually changes to fulfilled, then it would change the contract's status to fulfilled, given that all its obligations are fulfilled. 
\end{example}

\subsection{Contracts in ACT and GDPR compliance} \label{sec:gdpr-act}
\begin{figure}[!ht]
    \centering
    \begin{subfigure}[b]{1.00\textwidth}
        \centering
        \includegraphics[width=\linewidth]{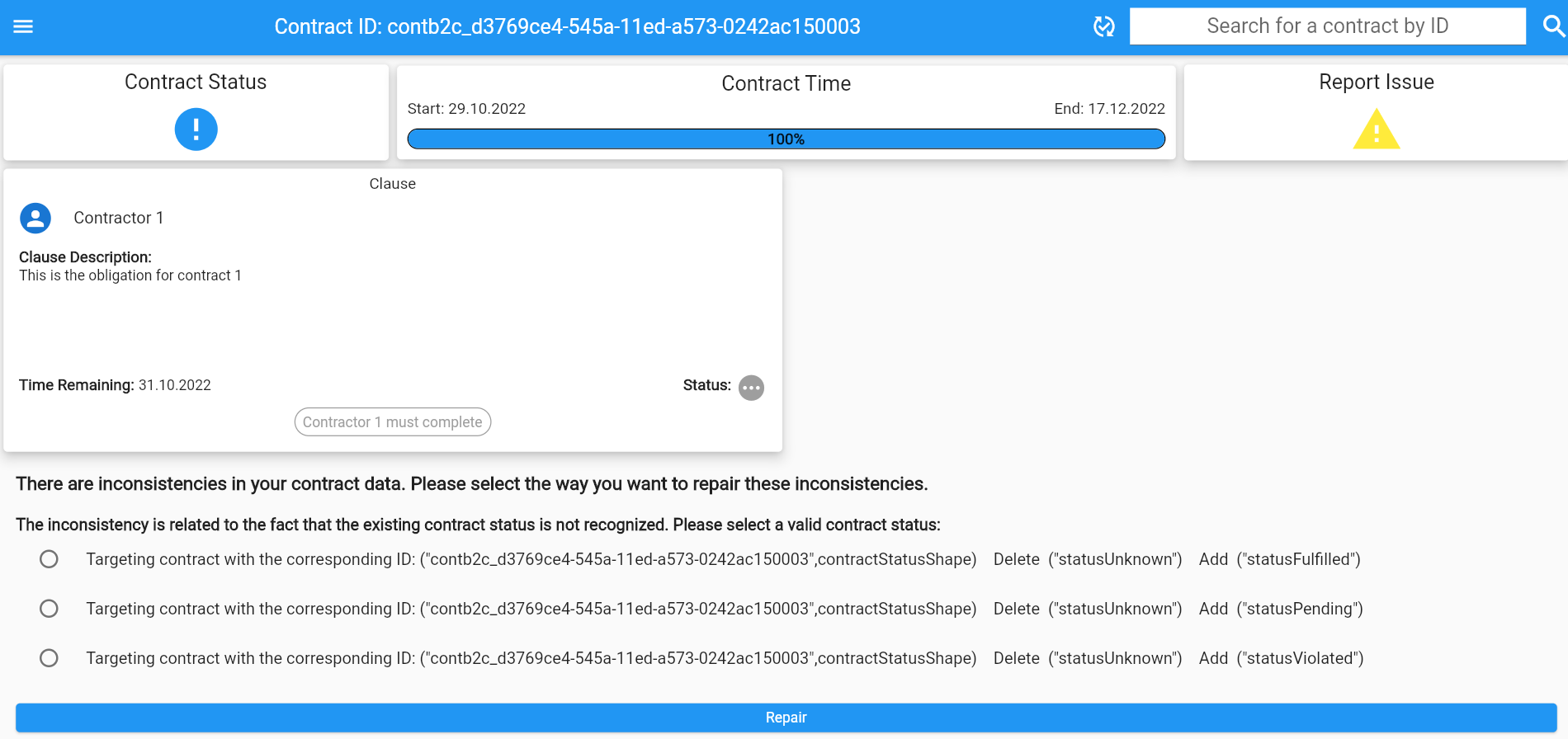}
        \label{fig:sub1}
    \end{subfigure}
    \hfill
    \begin{subfigure}[b]{1.00\textwidth}
        \centering
        \includegraphics[width=\linewidth]{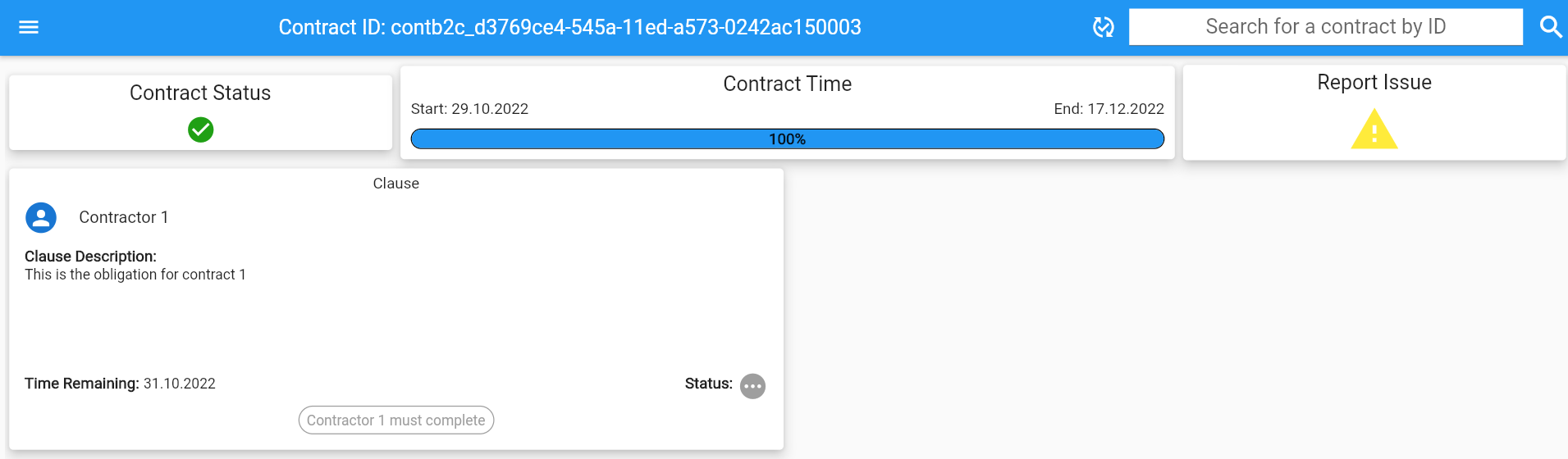}
        \label{fig:sub2}
    \end{subfigure}
    \caption{Repair prototype for ACT}
\label{fig:act-repairs}
\end{figure}

ACT is an application that is used to create and manage contracts between different parties along the contract lifecycle. Using its interface, one can associate clauses and obligations with contracts. The ACT UI flags whenever there is an obligation that has not been fulfilled in the respective due time by a party, i.e. GPDR compliance has been violated. In the backend, the contract data is stored in a triple store and accessed via SPARQL. ACT has been extended\footnote{\url{https://github.com/GeniBushati/cmt}} (cf. Fig.~\ref{fig:act-repairs}) to display whenever there are data inconsistencies and offers to restore consistency by choosing one of the options via human-in-the-loop. The repairs are devised from \emph{CCV Consistency Requirements} and \emph{SHACL Constraints} elaborated in the next section. 

Figure \ref{fig:act-repairs} shows the prototypical repair functionality implemented on top of ACT. The ``exclamation'' symbol (above) depicts that there is an inconsistency on the data level, which needs to be restored via applying repairs. The user can select one of the repairs (i.e. models) manually from the set of choices that are returned by the ASP program. The ``green check'' symbol (below) displayed after the consistency has been restored, i.e., after the user has opted for a repair.

\section{CCV Consistency and Repairs}\label{sec:ccv-constraints}
In this section, we introduce repair strategies to automatically ensure that every contract has a clearly defined and consistent status regarding the contract lifecycle. 
We first elaborate on formal consistency requirements (CR) of CCV and define them using logical expressions.
We focus on simple CRs for the states of contract and obligation of the semantic model to explain our approach. We see these CRs also as the most basic requirements regarding robustness from the distributed systems point of view. We then introduce SHACL constraints used by ACT to formally represent these requirements. 
We use the SHACL repair program to determine repair models to fix the constraint violations. However, now we aim to automatically determine one (optimal) repair choice to minimize the human-in-the-loop and provide self-correcting functionality. 
Therefore, we introduce repair strategies, which formalise optimzations to determine an optimal repair choice. 

\subsection{CCV Consistency Requirements}%
The CCV consistency requirements were originally determined for the contract data in the smashHit KG, which was built for the use cases of the smashHit project. 
They mainly address functional requirements for the lifecycle status, which needs to be exactly one status in every stage of the contract lifecycle.
Also, the state of obligations can affect the status of a related contract.
The CCV consistency requirements pose new challenges as they require to capture the expressivity of the \emph{logic implication} using SHACL shapes. 
In the following, each consistency requirement is explained and captured using formal logic.
We present three examples for CCV consistency requirements, which are selected for scope and simplicity reasons.\footnote{We note that further (and more complex) consistency requirements exist as  well for CCV, which can be incorporated using the same principles we outline herein. E.g., \emph{``A contract status is set to fulfilled if and only if all its obligations statuses are set to fulfilled''} boils down to the logical implication construct.}. 

\paragraph{CR-1}\label{req-1} The first consistency requirement is about the functional contract status and its values: 
\emph{Contracts must have exactly one contract status of pending, fulfilled or violated}.
\small
\begin{align*}
\forall x\exists y,z.&~Contract(x)~\wedge hasContractStatus(x, y) \wedge hasContractStatus(x, z)\\ &\wedge (y \neq z \lor
(y \neq \text{pending}~\wedge~y \neq \text{fulfilled}~\wedge~y \neq \text{violated}))\implies \bot
\end{align*}
\normalsize
\paragraph{CR-2}\label{req-2} The second consistency requirement is addressing the consistency of end dates of contract and obligations: \emph{Obligations must have exactly one end date and it must not exceed the end date of the associated Contract}.
\small
\begin{align*}
\forall x\exists y,z,w,v.&~Contract(x) \wedge Obligation(y) \wedge hasObligations(x,y) \\ &\wedge hasEndDate(x,z) \wedge hasEndDate(y,w)  \wedge hasEndDate(y,v) \\&\wedge ((w>z) \lor (w \neq v)) \implies \bot
\end{align*}
\normalsize

\paragraph{CR-3}\label{req-3} The third consistency requirement is about violating a contract: \emph{A Contract is violated if at least one associated Obligation has the state set to violated}.
\small
\begin{align*}
 \forall x\exists y,z.&~Contract(x) \wedge Obligation(y) \wedge hasObligations(x,y) \\ &\wedge hasState(y,z) \wedge z=\text{violated} \implies hasContractStatus(x, \text{violated})
\end{align*}
\normalsize

\subsection{SHACL Constraints}
In the following, we will define SHACL shapes which use constraint components to represent the consistency requirements CR-1, CR-2 and CR-3.
This translation of consistency requirements to SHACL is done using SHACL Core.
\paragraph{Functional Contract Status} The first shape implements the functionality and value requirements of a \emph{Contract}, as described in CR-1. 
\begin{lstlisting}
:FunctionalContractStatusShape a sh:NodeShape;
  sh:targetClass fibo:Contract;
  sh:property [
    sh:path smashHitCore:hasContractStatus;
    sh:in ( smashHitCore:statusPending smashHitCore:statusFulfilled smashHitCore:statusViolated );
    sh:minCount 1;
    sh:maxCount 1; ] .
\end{lstlisting}
\paragraph{Contract End Date} The second shape implements the dependency requirement of a contract regarding any associated obligation, where the end date of an obligation must not be higher than the end date of the contract, as described in CR-2.
\begin{lstlisting}
:EndDateConsistencyShape a sh:NodeShape;
    sh:targetClass smashHitCore:Obligation;
    sh:property [
        sh:path ( [ sh:inversePath smashHitCore:hasObligations] smashHitCore:hasEndDate);
        sh:lessThanOrEquals smashHitCore:hasEndDate 
 ];
    sh:property [
        sh:path smashHitCore:hasEndDate;
        sh:minCount 1;
        sh:maxCount 1; ] .
\end{lstlisting}
\paragraph{Contract Violation} The third shape implements the dependency requirement of a contract regarding any associated obligation, where one violated obligation leads to a violated contract, as described in CR-3. SHACL shape implementation in SHACL Core has been done using \emph{negation} and \emph{or} operators in order to model the \emph{logical implication}. 
\begin{lstlisting}
:ContractViolationShape a sh:NodeShape;
  sh:targetClass fibo:Contract;
  sh:or (   
    [ sh:not [
      sh:property [
        sh:path ( smashHitCore:hasObligations smashHitCore:hasState );
        sh:hasValue smashHitCore:ViolatedState ]; ] ]
    [ sh:property [
      sh:path smashHitCore:hasContractStatus;
      sh:hasValue smashHitCore:statusViolated; ] ] ).
\end{lstlisting}

\subsection{Repairing CCV Constraints Automatically}
Having SHACL shapes in place to implement the consistency requirements, the next step is to resolve inconsistencies when they occur.
When using CCV as part of an open environment, where different applications like ACT participate, 
we can take advantage of the repair approach to ensure CCV consistency on the data level.
For this scenario, we implement a self-correcting system that is robust against programming errors of participating applications by repairing CCV inconsistencies automatically, as an improvement from the GUI interface presented in \ref{sec:gdpr-act} where the choice is done by human-in-the-loop. 
For this implementation, we use the SHACL repair program as a basis to deduce repairs for the CCV data. 
These repairs, which are sets of additions and deletions, can then be applied to the graph data to achieve consistency regarding the constraints. 

\paragraph{Introducing Repair Strategies}
There is one missing step from deducing repairs to automatic repairs. In the case the SHACL repair program deduces multiple different options to repair the data, it will return all these possible repairs. 
However, we want to reduce these options and provide an optimal choice so that it is easier for the user to decide how to repair the data. 
Therefore, we discuss strategies for repair selection of the previously defined SHACL shapes, which then enable us to automatically present optimal repairs in the case of multiple options. 
We note that this practical work does not cover the formal semantics definition for repair strategies. 

\paragraph{Defining Repair Strategies}
To apply repair strategies to SHACL repairs, we developed a functional prototype to expand the existing SHACL repair processor with additional optimization rules, which are constraints and preferences. 
Optimizations will be added to the ASP repair program. However, we decided on a solution that makes it easier for users to define optimizations and that is close to the SHACL vocabulary. 
To define repair strategies, we introduce a new RDF-based vocabulary with a namespace \url{https://www.w3.org/ns/shacl/repairs}, abbreviated \emph{shr}, which includes the class \emph{shr:RepairStrategy}, the properties \emph{shr:hasConstraint}, \emph{shr:hasPreference}, \emph{shr:action}, \emph{shr:values}, \emph{shr:preferenceType}, \emph{shr:function} 
and the constants \emph{shr:add}, \emph{shr:delete},  \emph{shr:read-only}, \emph{shr:change}, 
\emph{shr:minValue} and \emph{shr:maxValue}. We also reuse some existing SHACL elements. 
We define the semantics informally as follows:
\begin{itemize}
    \item \emph{shr:RepairStrategy} is a class used to define a repair strategy with one or several repair optimizations, which are defined using the properties  \emph{shr:hasConstraint} and \emph{shr:hasPreference};
    \item \emph{shr:hasConstraint} defines a constraint for a given \emph{sh:path} and a given \emph{shr:action} property with value \emph{shr:add} or \emph{shr:delete}. There are two options to define which values are not allowed:
    \begin{itemize}
        \item If no values are given, then all value nodes of the given path are not allowed to be added or deleted, respectively; 
        \item The values are defined using the \emph{shr:values} property and a collection of values.
        \end{itemize}
    \item \emph{shr:hasPreference} defines a preference order for a given \emph{sh:path} and a given \emph{shr:action} property with value \emph{shr:add} or \emph{shr:delete}. \\
    The \emph{shr:preferenceType} determines if we prefer to keep (not add or delete) the data (value \emph{shr:read-only}) or change (add or delete) the data (value \emph{shr:change}). \\
    There are three options to define preferred values:
    \begin{itemize}
        \item If no values are given, then all value nodes of the given path are preferred to be added or deleted, respectively; 
        \item The values are listed in order using the \emph{shr:values} property and a list of values. 
        \item The values are determined from existing values in the graph based on a function given using \emph{shr:function} property and constants \emph{shr:minValue} or \emph{shr:maxValue};
        \end{itemize}
\end{itemize}
We implemented a repair strategy program on top of the SHACL repair program. It parses the RDF-based repair strategies and automatically generates additional rules to modify the ASP repair program to automatically choose the best option based on the defined strategies. 
These rules are clingo optimization rules, which can be used to determine optimal answer sets using weights, and additional constraints. 
%
%
In the following, we provide repair strategies for the CRs and show the logical rules in ASP that extend the repair program with optimization rules and constraints to implement these strategies.
\paragraph{Repairing CR-1}
CR-1 needs to be repaired by determining a single (unique) \emph{hasContractStatus} property atom with a value of \emph{pending}, \emph{fulfilled} or \emph{violated} to keep and removing the other \emph{hasContractStatus} atoms. 
We will pick one status based on the following preference order. 
If a contract or obligation is violated, we should keep this status and delete all other status. If a contract or obligation is fulfilled, but not violated, then we should keep fulfilled. In other cases the state is pending.
If there is no state, we choose to add the pending state to indicate that the contract or obligation is still open to be processed.
We formalize this repair strategy, which consists of two preferences. 
\begin{lstlisting}
:FunctionalContractStatusStrategy 
  a shr:RepairStrategy;
  shr:hasPreference [
    sh:path smashHitCore:hasContractStatus; 
    shr:action shr:delete;
    shr:preferenceType shr:read-only;
    sh:values ( smashHitCore:statusPending smashHitCore:statusFulfilled smashHitCore:statusViolated ) ];
  shr:hasPreference [
    sh:path smashHitCore:hasContractStatus; 
    shr:action shr:add;
    shr:preferenceType shr:change;
    shr:values ( smashHitCore:statusPending ) ] .
\end{lstlisting}
The first preference for deletions will determine a single model based on weights from the 3 possible repair models (one option for each of the 3 possible values). We define the preferred values using \emph{shr:values} with increasing weights based on this order. 
The actual weights must be in total order to be used for an unambiguous preference. In our case, the implementation uses $1,2,3, \dots$.
The option for \emph{statusPending} gets the lowest weight, followed by \emph{statusFulfilled} and finally \emph{statusViolated}. 
The weights enable the repair program to make a decision for the optimal choice by adding the following optimization rules to the repair program:
\small
\begin{align*}
& \#minimize { 1@0,X: del(hasContractStatus(X,statusPending)) } .\\
& \#minimize { 2@0,X: del(hasContractStatus(X,statusFulfilled)) } .\\
& \#minimize { 3@0,X: del(hasContractStatus(X,statusViolated)) } .
\end{align*}
\normalsize
These optimization rules implement the repair for deletions by prioritizing the values based on the defined order. They apply the weights to the possible repair models, thereby picking them for deletion in increasing order of the weight (first 1, then 2, then 3). 
Using this semantics, we can ensure that the preference is to not remove \emph{statusViolated}, whereas \emph{statusFulfilled} is only removed if there is a \emph{statusViolated}.

The second preference for additions will determine a preferred option to be added when there is not yet a status value. In this case, we maximize picking the value \emph{statusPending}, which is the preferred value, by assigning a weight of 1, while other values do not get a weight assigned.
We add the following optimization rule to the repair program:
\small
\begin{align*}
& \#maximize { 1@0,X: add(hasContractStatus(X,statusPending)) } .
\end{align*}
\normalsize
This optimization rule will maximize picking the addition of the \emph{statusPending}, meaning it will add this value as a preferred option over other values. 
\begin{example}
\label{example-r-1}
We assume the following data graph:
\begin{lstlisting}
:contb2b a fibo:Contract;
  smashHitCore:hasContractStatus smashHitCore:statusViolated, 
    smashHitCore:statusPending, smashHitCore:statusFulfilled .
\end{lstlisting}
This data graph has 3 different values for \emph{hasContractStatus}, which clearly violates the \emph{FunctionalContractStatusShape} and we get the 3 repair models with deletion sets $D_1$, $D_2$ and $D_3$.
\small
\begin{align*}
    D_1 = \{&hasContractStatus(contb2b,statusViolated),\\&hasContractStatus(contb2b,statusPending)\} \\
    D_2 = \{&hasContractStatus(contb2b,statusPending),\\&hasContractStatus(contb2b,statusFulfilled)\} \\
    D_3 = \{&hasContractStatus(contb2b,statusViolated),\\&hasContractStatus(contb2b,statusFulfilled)\}
\end{align*}
\normalsize
When we apply the repair strategy to it, the application will prioritise the models based on the weights for the different values, resulting in a single optimal model:
\small
\begin{align*}
    D_2 = \{&hasContractStatus(contb2b,statusPending),\\&hasContractStatus(contb2b,statusFulfilled)\}
\end{align*}
\normalsize
$D_2$ has the lowest weight of the 3 models and is therefore picked by the optimization rules as the single optimal repair.
\end{example}

\paragraph{Repairing CR-2}
CR-2 states that the value of \emph{hasEndDate} of an obligation must not be higher than the end date of the associated contract. 
The repair program will remove any \emph{hasEndDate} atoms from the obligations with a value that is not less than or equal to the end date of the contract, and it adds a \emph{hasEndDate} atom with the end date of the contract if the obligation misses a lower or equal end date.
However, the repair program might still return multiple models in the case the obligation already has multiple end dates which are lower or equal to the end date of the contract. In this case we want the repair strategy to pick one of these end dates as a preferred end date. 
We decide for a strategy to keep the maximum value, which is the highest end date and as such the closest one to the end date of the contract. The repair strategy proposes this option to the user as the safest approach to repair the data from a real-world perspective. 
We define a preference to keep (\emph{shr:read-only}) the maximum value(s) (\emph{shr:maxValue}), which means that smaller values are deleted first.
\begin{lstlisting}
:FunctionalEndDateStrategy a shr:RepairStrategy;
    shr:hasPreference [
        sh:path smashHitCore:hasEndDate; 
        shr:action shr:delete;
        shr:preferenceType shr:read-only;
        shr:function shr:maxValue ] .
\end{lstlisting}
The following optimization rule is added to the repair program:
\small
\begin{align*}
\#minimize \{ &1@0,X,Y: del(hasEndDate(X,Y)),\\
&hasEndDate(X,Z,t^{**}), Y>=Z \} .
\end{align*}
\normalsize
This optimization rule will minimize picking the deletions of the largest end date of the \emph{hasEndDate} atoms, meaning it will keep the latest date as a preferred option over other values.
\begin{example}
\label{example-r-2}
We assume the following data graph:
\begin{lstlisting}
:contb2b a fibo:Contract;
  smashHitCore:hasEndDate "2022-09-07"^^xsd:dateTime; 
  smashHitCore:hasObligations :ob_1 .
:ob_1 a smashHitCore:Obligation; 
  smashHitCore:hasEndDate "2021-09-07"^^xsd:dateTime, "2020-09-07"^^xsd:dateTime .
\end{lstlisting}
Without the repair strategy, we get the 2 repair models with deletion sets $D_1$ and $D_2$.
\small
\begin{align*}
    D_1 = \{&hasEndDate(ob\_1,"2021-09-07")\} \\
    D_2 = \{&hasEndDate(ob\_1,"2020-09-07")\}
\end{align*}
\normalsize
However, when we apply the repair strategy, the repair program will pick the repair with the deletion for the lower end date for the obligation, resulting in a single optimal model with the highest end date:
\small
\begin{align*}
    D_2 = \{&hasEndDate(ob\_1,"2020-09-07")\}
\end{align*}
\normalsize
\end{example}

\paragraph{Repairing CR-3}
CR-3 states that a contract is violated if at least one obligation is violated. 
In terms of repair strategy this means that there must not be any non-violated contract if there is a violated obligation. 
In case of a violation, the repair program either deletes \emph{hasObligations}, removes the \emph{ViolatedState} from the obligations or it adds the \emph{statusViolated} to the contract. 
The repair strategy we choose to implement the CCV semantics is to only allow adding \emph{statusViolated} for \emph{hasContractStatus} if there is an obligation with \emph{ViolatedState}. 
We thereby infer the contract status based on the obligation states in case of violations. 
We define two read-only constraints for the repair strategy. 

%
%
\begin{lstlisting}
:ViolatedContractStrategy a shr:RepairStrategy;
  shr:hasConstraint [
    sh:path smashHitCore:hasObligations; 
    sh:action sh:delete; ];
  shr:hasConstraint [
    sh:path smashHitCore:hasState; 
    shr:action shr:delete;
    shr:values ( smashHitCore:ViolatedState; ) ] .
\end{lstlisting}
The following two constraints are added to the repair program based on the two read-only constraints in the repair strategy:
\small
\begin{align*}
& :-del(hasObligations(X,Y)) .
& :-del(hasState(X,ViolatedState)) .
\end{align*}
\normalsize
The first read-only constraint prevents the deletion of any \emph{hasObligations} properties, which means disconnecting the contract from the obligation is no longer an available repair choice. 
The second read-only constraint prevents picking the deletion of \emph{ViolatedState} for obligations as a repair choice. 
Specifically, the constraints prevents models with these deletions at all before any optimizations are done. 
The remaining repair choice is the addition of \emph{statusViolated} as a value for \emph{hasContractStatus}. 
\begin{example}
\label{example-r-3}
We assume the following data graph:
\begin{lstlisting}
:contb2b a fibo:Contract;
  smashHitCore:hasContractStatus smashHitCore:statusFulfilled; 
  smashHitCore:hasObligations :ob_1 .
:ob_1 a smashHitCore:Obligation; 
  smashHitCore:hasState smashHitCore:ViolatedState .
\end{lstlisting}
Without the repair strategy, the repair program would return the following two minimal models with deletions $D_1$ and $D_2$
\small
\begin{align*}
    D_1 = \{hasObligations(contb2b,ob\_1)\} \qquad
    D_2 = \{hasState(ob\_1,ViolatedState)\}
\end{align*}
\normalsize
These minimal repairs would both violate the CCV semantics.
However, when we apply the repair strategy, the repair program will prevent models to pick \emph{hasObligations} or \emph{ViolatedState} to be deleted, thereby preventing these minimal models to be picked. The consequence is a new minimal model under the constraints added by the repair strategy, which has additions $A_1$ and deletions $D_1$:
\small
\begin{align*}
    &A_1 = \{hasContractStatus(contb2b,statusViolated)\}\\
    &D_1 = \{hasContractStatus(contb2b,statusFulfilled)\}
\end{align*}
\normalsize
\end{example}
With repair strategies, the data can now only be repaired when assigning the contract status \emph{statusViolated} to the contract, thereby implementing the expected semantics regarding data consistency.

\section{Evaluation of CCV Repairs}\label{sec:evaluation}
%
%
%
%
%
Our motivation for repair strategies is to propose optimal repair options to users within the ACT tool to fix inconsistent CCV data. 
To verify our technical implementation, we defined test scenarios based on (anonymous) contract data from the smashHit KG, which includes scenarios from the use cases of the smashHit project, and performed evaluations regarding correctness and performance. 
First, we implemented repair strategies to extend SHACL repairs and then defined unit tests to cover the previously described consistency requirements regarding correctness. 
Second, we applied bulk performance tests to cover the scenario where an unprecedented amount of data needs to be simultaneously processed by multiple agents. 

\paragraph{Bulk Tests}
The bulk tests are based on contract and obligation data from the test case scenarios for ACT. We aim to simulate large scale data inconsistencies by generating data to violate the SHACL constraints using the following approach. 
The basis is a consistent snapshot of existing contract and obligation data, which includes 19 contracts and 21 obligations (1374 triples total). 
For these contracts and obligations, we randomly generate additional status and end dates to cause inconsistencies. The randomization is done regarding how many to add (up to 4) in addition to the existing status or end date. We also randomly generate additional states for the obligations, where at least one violated obligation \emph{ViolatedState}) must lead to a violated contract (\emph{statusViolated}) as well. 
Using this approach, we generated a total number of 200 test cases with random inconsistencies ranging from 0 to up to 3000. 
We note that we use a 3-threaded customized clingo configuration, which we previously determined in experiments to be the best options regarding performance in our scenario. 
We define the correctness as the acceptance criteria for the simulated tests as the following:
\begin{itemize}
    \item All the constraints are repaired, resulting in a consistent data graph.
    \item There is one (optimal) repair model returned.
\end{itemize}
We also measure the performance of the test runs and analyze it afterwards.

\paragraph{Test Results}
All 200 tests were completed successfully with returning a single repair model representing an optimal choice, thereby satisfying the defined acceptance criteria. 
For discussing the performance of the CCV repairs, we generated a plot (below) of the test performance data in relation to the number of inconsistencies.
We can see that up to 3000 inconsistencies, the data was repaired within a minute. 
The performance curve shows an above linear development. As a conclusion from the test results, our approach shows promising performance in practice, highlighting the approach's correctness and scalability. 
The implementation of the repair strategies, the unit tests and the real-world contract data for the bulk tests can be viewed at GitHub\footnote{\url{https://github.com/robert-david/shacl-repairs/tree/ccv-repair-strategies}}.
\begin{center}
\begin{tikzpicture}
\begin{axis}[width=235,
    height=\axisdefaultheight-80,
    xlabel = {Data graph inconsistencies},
    ylabel = {Solving time (sec.)}
    ],
\addplot[only marks,mark=*,mark options={scale=0.3}
] file {figures/ccv-evaluation.txt};
  \end{axis}
\end{tikzpicture}
\end{center}

\section{Conclusions \& Future Work}\label{sec:Conclusion}
%
%
In this work, we presented a novel approach for semi-automatically resolving data inconsistencies for CCV data managed by the ACT tool. 
We implemented a repair strategy approach which reduces the number of repair choices to optimal choices. 
Presenting these choices to users of the ACT tool makes it easier to decide how to repair inconsistencies and thereby support them in managing GDPR-compliant contract lifecycle data. 
Although the constraints presented here are rather a simple first step in this application, it is easy to expand on these and cover more complex contract scenarios, because the implementation is based on the standardized SHACL Core language and can be automatically translated into repairs. 
Therefore, as a next step extending our problem setting, we will investigate the further challenges that occur in an extended problem definition. For instance, by extending the problem space to include a number of relations (e.g., \textit{Contractors} via \textit{smashHitCore:hasContractors}) that require property paths for constraint validation and accommodated repairs. 
Regarding the quality of the generated repairs, we aim to assess the overlap between expert user decisions for repair and the outcomes from the automatic repair strategies in order to evaluate which preferences of repair choices make most sense. 
Furthermore, we will look into further use cases that can be covered by SHACL and investigate if the repair strategy approach is generally applicable to a wider range of repair scenarios. 
Addressing the foundation for this approach on the theoretical side, we aim to provide a formal semantics for repair strategies. 

In a broader context, our work on contract repairs has a direct potential impact on the adoption and evolution of semantic access and usage control mechanisms by enhancing the practical applicability of semantic contract environments. Specifically, our developments contribute to the robustness and reliability of such environments, thereby facilitating the adoption and effective utilization of ODRL and similar vocabularies for expressing, validating, and enforcing contractual agreements. By addressing inconsistencies and ambiguities in contract specifications, our approach ensures that semantic contracts remain both machine-processable and legally interpretable, reinforcing the role of policy languages in digital governance and rights management. Moreover, our work is likely to have a direct impact on the ODRL and the like languages themselves, as ensuring the quality of contractual data and systematically representing knowledge about its correctness and integrity in semantic annotations are essential for the evolution and refinement of policy languages. Given the increasing reliance on formalized contracts in automated systems, integrating contract repair mechanisms is not only a valuable enhancement but also a necessary step towards the broader adoption and scalability of semantic policy frameworks. Therefore, advancing methodologies for contract repair represents a strategic direction for realizing the full potential of ODRL and beyond in real-world applications.
%
\section*{Declaration on Generative AI}
The author(s) have not employed any Generative AI tools.

\bibliography{bibliography}

\end{document}